\documentclass[12pt]{article}
\usepackage{hyperref}
\usepackage{cite}
\usepackage{color}
\usepackage{graphicx}

\def\baselinestretch{1.2}
\newcommand{\be}{\begin{equation}}
\newcommand{\ee}{\end{equation}}
\newcommand{\beq}{\begin{eqnarray}}
\newcommand{\eeq}{\end{eqnarray}}

\newcommand{\gone}[1]{{}}

\begin{document}
\begin{titlepage}
\begin{flushright}
MAD-TH-08-12
\end{flushright}

\vfil

\begin{center}

{\Large{\bf Supergravity dual of Chern-Simons Yang-Mills theory with ${\cal N}=6,8$ superconformal  IR fixed point}}

\vfil

Akikazu Hashimoto and Peter Ouyang

\vfil

Department of Physics, University of Wisconsin, Madison, WI 53706

\vfil

\end{center}

\begin{abstract}
\noindent We construct a solution of eleven dimensional supergravity
corresponding to a stack of M2 branes localized at the center of a
particular eight dimensional hyper-K\"ahler manifold constructed by
Gauntlett, Gibbons, Papadopoulos, and Townsend, generalizing the
earlier construction of Cherkis and Hashimoto. In the decoupling
limit, this solution is dual to a Chern-Simons/Yang-Mills/Matter
theory in 2+1 dimensions with ${\cal N}=3$ supersymmetry, which
flows in the infra red to a superconformal Chern-Simons/Matter
system preserving ${\cal N}=6,8$ supersymmetry, constructed recently
by Aharony, Bergman, Jafferis, and Maldacena.
\end{abstract}
\vspace{0.5in}

\end{titlepage}
\renewcommand{\baselinestretch}{1.05}  

Until recently, there was no known formulation of superconformal
Chern-Simons theory in 2+1 dimensions with ${\cal N}=8$ supersymmetry,
and in fact the theory was believed not to exist
\cite{Schwarz:2004yj}. This belief was reversed by the explicit
construction of a model with ${\cal N}=8$ supersymmetry by Bagger
and Lambert \cite{Bagger:2007jr,Gustavsson:2007vu}.  The original
formulation of Bagger, Lambert, and Gustavsson involved the use of a
3-algebra, of which only a single finite dimensional example with a
positive definite metric, the $SO(4)$ 3-algebra, is known to exist
\cite{Papadopoulos:2008sk,Gauntlett:2008uf}.  Shortly after its
construction, this $SO(4)$ model was shown to be equivalent to a more
traditional Chern-Simons theory with an $SU(2) \times SU(2)$ gauge
group, and matter fields in the bi-fundamental representation
\cite{VanRaamsdonk:2008ft,Bandres:2008vf} with does not rely on the
use of a 3-algebra. These theories are extremely interesting as a
candidate Lagrangian description of the decoupled field theory of
M-theory membranes. In the past several months, there has been
significant progress in the understanding of this model and its
generalizations reported in the literature.

\begin{figure}
\centerline{\includegraphics[width=2.5in]{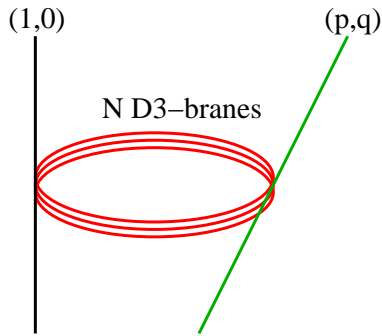}}
\caption{A configuration of D3, NS5, and $(p,q)$ 5-branes in type IIB
string theory. $N$ D3-branes wind around an $S_1$ of size $L$. An
NS5-brane and a $(p,q)$ 5-brane intersects the D3-brane at a localized
point along the $S_1$ but extends along the other 3 world volume
coordinates of the D3-branes.  Low energy effective theory of open
strings is a Chern-Simons/Yang-Mills/matter theory with gauge group
$U(N) \times U(N)$.
\label{figa}}
\end{figure}

A very interesting new perspective on these class of models from the
point of view of string theory was recently presented by Aharony
et.al.  \cite{Aharony:2008ug}. These authors considered a
configuration of branes in type IIB string theory involving D3-branes,
NS5-branes, and $(p,q)$ 5-branes of the form illustrated in figure
\ref{figa}.  By $(p,q)$ 5-brane, we mean the bound state of $p$
NS5-branes and $q$ D5-branes. More specifically, we orient the
D3-branes along the 0126 directions. We take the 6 direction to be
compact. The NS5-branes are oriented along the 012345 directions,
and the $(p,q)$ 5-branes are oriented along the $012[{3 ,
7}]_{\theta} [{4 , 8}]_{\theta} [{5 , 9}]_{\theta}$ directions. We
are following the notational conventions of \cite{Aharony:2008ug}.
This brane configuration is a particular case of class of
configurations considered in \cite{Kitao:1998mf,Bergman:1999na}
which generalizes the construction of Hanany-Witten type
\cite{Hanany:1996ie}.  Localized intersection of $(p,q)$ 5-brane and
D3-brane was also studied recently in \cite{Gaiotto:2008sd}.  If
$(p,q) = (1,0)$, we recognize this system as describing an  impurity
system \cite{Sethi:1997zza,Kapustin:1998pb} in 3+1 dimensions
\cite{Karch:2001cw,DeWolfe:2001pq} which flows to a 2+1 dimensional
$U(N) \times U(N)$ Yang-Mills theory with bi-fundamental matter
preserving ${\cal N}=4$ supersymmetry.  For $(p,q)=(1,k)$, one also
obtains a defect field theory which flows to a $U(N) \times U(N)$
Yang-Mills theory with a Chern-Simons level $k$ and matter in the
bi-fundamental representations. These configurations generically
preserve ${\cal N}=3$ supersymmetry
\cite{Kitao:1998mf,Bergman:1999na}. The main observations of
\cite{Aharony:2008ug} are as follows:
\begin{itemize}
\item The level $k$ $U(N) \times U(N)$ Chern-Simons/Yang-Mills/matter
theory flows in the IR to a level $k$ $U(N) \times U(N)$
Chern-Simons/matter theory with no Yang-Mills kinetic term.

\item For $k> 2$, the IR theory has ${\cal N}=6$ superconformal symmetry

\item For $k=1$ and $k=2$, the supersymmetry of the IR theory is
enhanced to ${\cal N}=8$.
\end{itemize}
Aharony et.al.\ also noted that had they considered the gauge group
$SU(2) \times SU(2)$, this model is equivalent to the product gauge
group formulation \cite{VanRaamsdonk:2008ft,Bandres:2008vf} of the
Bagger-Lambert-Gustavsson theory.  From this point of view, the role
of the 3-algebra is demoted to the coincidence of the structure of the
$SU(2) \times SU(2)$ group, while the brane construction provides a
plethora of models with ${\cal N} \ge 6$ supersymmetry where the
features such as the product gauge group and the bi-fundametal matter
content have natural origins.\footnote{A formulation of ${\cal N}=6$
theory in terms of 3-algebra appeared in a recent article
\cite{Bagger:2008se}}. The formulation of \cite{Aharony:2008ug} also
identifies the $U(N) \times U(N)$ at $k=1$ as the candidate
Lagrangian description for the stack of $N$ M2-branes.
Unfortunately, when $k=1$ the model is strongly coupled, making the
analysis of interesting features such as the $N^{3/2}$ scaling of
the entropy beyond reach for the time being.

In order to see the enhancement of supersymmetry from ${\cal N}=3$
to ${\cal N}=6$ or 8 \cite{Aharony:2008ug}, it is useful to
T-dualize the configuration of figure \ref{figa} along the 6
coordinate and lift to M-theory. This gives rise to a configuration
of $N$ M2-branes in eleven dimensions, compactified in 2 cycles, the
(6,11), transverse to the world volume of the M2. The $(1,0)$ and
the $(p,q)$ 5-branes are mapped to an overlapping configuration of
KK5-branes with charged $(1,0)$ and $(p,q)$ with respect to the
$U(1)\times U(1)$ associated with the 6 and 11 cycles, respectively.
As it turns out, the complete supergravity description of these
overlapping KK5-branes is known from the work of
\cite{Gauntlett:1997pk}.  It can be described as an eight
dimensional geometry with $sp(2)$ holonomy, and for general $(p,q)$
gives a family of geometries generalizing Taub-NUT $\times$ Taub-NUT
space which has holonomy group $sp(1) \times sp(1)$. With the
$sp(2)$ holonomy, the geometry is hyper-K\"ahler and preserves
$3/16$ of the supersymmetries of the eleven dimensional
supergravity, which is precisely what we expect for the dual of
theories in 2+1 dimensions with ${\cal N}=3$ supersymmetry.  These
spaces have also appeared as moduli-space of BPS monopoles
\cite{Gibbons:1995yw,Lee:1996kz}.  Just as in the case of the
Taub-NUT geometry, the overlapping KK5-brane has a core region which
is an orbifold $C^4 / Z_k$ where the discrete symmetry $Z_k$ rotates
each of the four complex plane in $C^4$ by an amount $2 \pi / k$.
Such an orbifold preserves $3/8$ of the supersymmetry of eleven
dimensional supergravity for $k>2$ and $1/2$ for $k=1,2$
\cite{Nilsson:1984bj}. Adding the M2 branes does not break any
further supersymmetries. For a theory with large $N$ and large 't
Hooft coupling $\lambda=N/k$, we are lead to take the gravitational
back reaction of the M2 branes into account, giving rise to a dual
$AdS_4\times S_7/Z_k$ geometry.

Let us now consider taking the limit where the cycle along the
6-direction, transverse to the M2-brane, is made arbitrarily large.
This amounts to making the compact world volume of the D3-brane along
the 6 direction in the original type IIB description, illustrated in
figure \ref{figa}, small. We would then have a
Chern-Simons/Yang-Mills/matter system on the world volume of the
D2-brane which we can decouple from gravity provided we scale the
radius along the eleven direction appropriately.

It is possible to consider the supergravity dual of this configuration
by taking the gravitational back reaction of the M2 branes into
account. Such a description would be appropriate for large
$N$. Finding the gravitational back reaction of the M2-brane amounts
to finding a solution to Laplace's equation with a source in the
background of the overlapping KK5-brane geometry. Once the Laplace
equation is solved, it is straight forward to embed it into the
solution to the equation of motion of eleven dimensional supergravity
using the standard ansatz.

In fact, a problem very similar to this was discussed for the case
where the KK5-brane geometry simplified to $R^4 \times
\mbox{Taub-NUT}$ or $\mbox{Taub-NUT} \times \mbox{Taub-NUT}$
\cite{Cherkis:2002ir} where the holonomy group is $sp(1)$, and $sp(1)
\times sp(1)$, respectively. The harmonic function is generically a
solution to linear, partial differential equation. In
\cite{Cherkis:2002ir}, the Laplace equation was solved using brute
force separation of variables. The resulting supergravity solution was
interpretable as being dual to 2+1 dimensional SYM with matter in the
fundamental representation. Regardless of the matter content,
Yang-Mills theory in 2+1 dimensional is superrenormalizable, and as
such, this supergravity solution is a dual of a UV complete field
theory.

The goal of this paper is to solve for the analogous harmonic
function for the overlapping KK5-brane geometry. By taking the
appropriate decoupling limit, we obtain a supergravity solution
which one can interpret as being dual to a specific
Chern-Simons/Yang-Mills/matter theory in 2+1 dimensions.  We will
examine the form and the tractability of the Laplace equation in
this background, with the expectation that the $sp(2)$ special
holonomy should provide some degree of analytic control. Note that
this precise program was outlined in the last paragraph of
\cite{Cherkis:2002ir}.

Let us begin by reviewing the basic ansatz for the intersecting brane
configuration following \cite{Cherkis:2002ir}.  We consider the ansatz
\beq ds^2  & = & H^{-2/3} (-dt^2 + dx_1^2+dx_2^2) + H^{1/3} ds_{{\cal M}8}^2 \\
F & = & dt \wedge dx_1 \wedge dx_2 \wedge dH^{-1} \label{ansatz} \eeq
where ${\cal M}_8$ is the eight dimensional $sp(2)$ holonomy manifold,
and $H(y_i)$ is a scalar function depending only on the coordinates of
${\cal M}_8$. By substituting this ansatz into the equation of motion
of supergravity in eleven dimensions, one can show that $H$ is
required to solve the Laplace equation in ${\cal M}_8$.

Next, let us review the metric for ${\cal M}_8$
\cite{Gauntlett:1997pk}. It is given by
\be ds^2 = V_{ij} d\vec y_i d\vec y_j + (V^{-1})^{ij} R_i R_j (d \varphi_i + A_i) (d \varphi_j + A_j) \ee
where
\be V_{ij} = \delta_{ij} + {1 \over 2} {R_i p_i R_j p_j \over |R_1 p_1 \vec y_1+ R_2 p_2 \vec y_2| }
+
{1 \over 2} {R_i \tilde p_i R_j \tilde p_j \over |R_1 \tilde p_1 \vec y_1+ R_2 \tilde p_2 \vec y_2| } \ ,
\ee
$i,j$ take values $1,2$, and $\vec y_i$ are 3 vectors. We have
restricted our attention to the case where there are two overlapping
KK5-branes whose charges are
\be (p_1,p_2) = (1,0), \qquad (\tilde p_1, \tilde p_2) = (1,k) \ee
to match the construction of \cite{Aharony:2008ug}. The $\varphi_i$
coordinate is chosen to have period $2 \pi$. So $R_1$ and $R_2$ are
radius of $S_1 \times S_1$ which we identify as the 6 and 11
directions, respectively. Therefore, when taking the decoupling limit,
we scale
\be R_1 = {2 \pi \alpha' \over L}, \qquad R_2 = g_s l_s = c g_{YM2}^2 \alpha' \label{scaling}\ee
where $c=(2 \pi)^{p-2}=1$ for $p=2$ \cite{Itzhaki:1998dd}, and $L$ is
the size of the circle along the 6-direction in the dual type IIB
description illustrated in figure \ref{figa}. We will eventually take
$L \rightarrow 0$, keeping $g_{YM2}^2$ fixed. This amounts to taking
the limit $R_2/R_1 \rightarrow 0$.

The simplest and the most symmetric case to consider is to place the
M2-brane at the origin $\vec y_1=\vec y_2=0$. We also restrict our
attention to a solution symmetric with respect to shifts in
$\varphi_1$ and $\varphi_2$. In the near core region, this is simply
the rotational symmetry of the ansatz.

It is then straight forward to write the Laplace equation on this
geometry as
\be 0= \partial_\mu (\sqrt{g} g^{\mu \nu } \partial_\nu H) =
\vec \partial_i \det V (V^{-1})^{ij} \vec \partial_j H(\vec y_1, \vec y_2) \ .\ee
This can be simplified a little by changing variables
\be \vec r_1 = \vec y_1, \qquad \vec r_2 = \vec y_2+{R_1 \over k  R_2} \vec y_1 \ . \ee
The Laplace equation will then have the form
\be
\left[\left(1  + {k R_2 \over 2 r_2}\right) \vec \partial_1^2
+{2 R_1 \over k R_2} \vec \partial_1 \cdot \vec \partial_2 +
\left(1 + {R_1^2 \over k^2 R_2^2} + {R_1 \over 2 r_1}\right) \vec \partial_2^2\right] H(\vec r_1, \vec r_2)  \ . \label{simp1}
\ee
%
The most symmetric configuration can depend, in general, on
\be r_1 = | \vec r_1|, \qquad r_2 = |\vec r_2|, \qquad z = {\vec r_1 \cdot \vec r_2 \over r_1 r_2} \ . \ee
In terms of these variables, the differential operators appearing in
\ref{simp1} have the form
\be \vec \partial_i^2 = {1 \over r_i} \left({\partial \over \partial  r_i} \right)^2 r_i + {1 \over r_i^2} \left( (1 - z^2) \partial_z^2 - 2 z \partial_z\right)  \label{lap1}\ee
\be  \vec \partial_1 \cdot \vec \partial_2 = {1 \over r_1 r_2}\left( z (z^2-1) \partial_z^2 + (1+z^2) \partial_z \right)  + {(1-z^2)  \over r_1} \partial_{r_2} \partial_z
+ {(1-z^2)  \over r_2} \partial_{r_1} \partial_z  + z \partial_{r_1} \partial_{r_2}\ .
\ee
At this point, we are faced with a linear yet seemingly unseparable
partial differential equation of three variables, with no obvious hope
for any simplification.

We are, however, entitled to take the large $R_1$ limit. To do this,
it is convenient to make the change of variables standard in taking
the near core limit of a Taub-NUT geometry
\be r_1 = {\rho_1^2 \over 2 R_1}, \qquad r_2 = {\rho_2^2 \over 2 k R_2} \ . \ee
In these coordinates, the metric on ${\cal M}_8$ has the form
\beq ds_{{\cal M}8}^2 &=&
\left(1+\left({1 \over R_1^2}+{1 \over k^2 R_2^2} \right) \rho_1^2\right)
\left(d \rho_1^2  + {\rho_1^2 \over 4} (d \theta_1^2 + \sin^2 \theta_1 d \phi_1^2) \right) \cr
&& + \left(1 + {\rho_2^2 \over k^2 R_2^2}\right)
\left(d \rho_2^2  + {\rho_2^2 \over 4} (d \theta_2^2 + \sin^2 \theta_2 d \phi_2^2) \right) \cr
&& -  {\rho_1^2 \rho_2^2 \over 2k^2 R_2^2} \left( {d \vec r_1 \cdot d \vec r_2 \over r_1 r_2} \right) \cr
&&+ \left( \begin{array}{cc}
{1 \over R_1^2} +  {1 \over \rho_1^2} + {1 \over  \rho_2^2} & {k \over  \rho_2^2} \\
{k \over \rho_2^2} & {1 \over R_2^2} + {k^2 \over \rho_2^2}\end{array}\right)^{-1\, ij} (d \varphi_i + A_i) (d \varphi_j + A_j)
\eeq
where $\theta_i$ and $\phi_i$ are the angular coordinates in $S_2$
of $\vec r$.  We have not reparameterized the term proportional to
$d \vec r_1 \cdot d \vec r_2 / r_1 r_2$ but it should be clear that
this expression is independent of $R_1$.  After taking $R_1
\rightarrow
\infty$ keeping $\rho_i$ and $R_2$ fixed, the harmonic equation
becomes
\beq 0 &=& \left[\left(1 +{\rho_2^2 \over  k^2 R_2^2}\right)
\left( \partial_{\rho_1}^2 + {3 \over \rho_1} \partial_{\rho_1} + { {4  \over \rho_1^2 } \left( (1 - z^2) \partial_z^2 - 2 z \partial_z\right) }\right) \right.\cr
&&+
\left(1 + {\rho_1^2 \over  k^2 R_2^2}\right)
 \left( \partial_{\rho_2}^2 + {3 \over \rho_2} \partial_{\rho_2} + { {4  \over \rho_2^2 } \left( (1 - z^2) \partial_z^2 - 2 z \partial_z\right) }\right)\cr
&& \left. +{ 2 \rho_1^2 \rho_2^2 \over  k^2 R_2^2 }
\left({\vec \partial_1 \cdot \vec \partial_2 \over k  R_1 R_2} \right)\right] H(\rho_1,\rho_2,z)
\eeq
where the expression
\beq  {\vec \partial_1 \cdot \vec \partial_2 \over k R_1 R_2}  &=& {4 \over \rho_1^2 \rho_2^2}\left( z (z^2-1) \partial_z^2 + (1+z^2) \partial_z \right)  \cr
&& + {2  \over \rho_1^2 \rho_2} (1-z^2) \partial_{\rho_2} \partial_z
+ {2  \over \rho_1 \rho_2^2} (1-z^2) \partial_{\rho_1} \partial_z  +
{ z \over \rho_1 \rho_2}  \partial_{\rho_1} \partial_{\rho_2} \label{rho2}
\eeq
is independent of $R_1$ despite appearances. Although this equation is
still not separable, we see that if $R_2 \rightarrow \infty$, this
equation simplifies to
\beq 0 &= &
 \left[
\left( \partial_{\rho_1}^2 + {3 \over \rho_1} \partial_{\rho_1} + {{4  \over \rho_1^2 } \left( (1 - z^2) \partial_z^2 - 2 z \partial_z\right) }\right) \right. \cr
&& \left. +
\left( \partial_{\rho_2}^2 + {3 \over \rho_2} \partial_{\rho_2} + { {4  \over \rho_2^2 } \left( (1 - z^2) \partial_z^2 - 2 z \partial_z\right) }\right) \right] H(\rho_1, \rho_2,z)
\eeq
which is separable. An obvious solution is
\be H_0 = {c'  N l_p^6 \over (\rho_1^2 + \rho_2^2)^3} \ , \label{seed}\ee
where $c' = 2^5 \pi^2$ \cite{Maldacena:1997re}. Such simplicity is
exactly what we expect since in the when $R_2 \rightarrow \infty$,
we are working in the near core limit where ${\cal M}_8 = C^4/Z_k$.

Let us now look at how the harmonic equation depends on $R_2$. One can
in fact collect its dependence on $R_2$ and write down a recursion
relation
\be {\bf A} H_{i+1}(\rho_1,\rho_2,z) = - {\bf B} H_i(\rho_1,\rho_2,z) \ee
where {\bf A} and {\bf B} are differential operators
\be {\bf A} =
\left( \partial_{\rho_1}^2 + {3 \over \rho_1} \partial_{\rho_1} + { {4  \over \rho_1^2 } \left( (1 - z^2) \partial_z^2 - 2 z \partial_z\right) }\right)
+
 \left( \partial_{\rho_2}^2 + {3 \over \rho_2} \partial_{\rho_2} + { {4  \over \rho_2^2 } \left( (1 - z^2) \partial_z^2 - 2 z \partial_z\right) }\right)  \ee
\beq {\bf B} & = & \left[{\rho_2^2 \over k^2 R_2^2} \left( \partial_{\rho_1}^2 + {3 \over \rho_1} \partial_{\rho_1}  +{ {4  \over \rho_1^2 } \left( (1 - z^2) \partial_z^2 - 2 z \partial_z\right) }\right)  \right. \cr
&&  \left. + {\rho_1^2 \over  k^2 R_2^2}
\left( \partial_{\rho_2}^2 + {3 \over \rho_2} \partial_{\rho_2} +{{4  \over \rho_2^2 } \left( (1 - z^2) \partial_z^2 - 2 z \partial_z\right) }  \right)
+
{ 2 \rho_1^2 \rho_2^2 \over  k^2 R_2^2 }
\left({\vec \partial_1 \cdot \vec \partial_2 \over k R_1 R_2}\right)\right] \ .
\eeq

This means
\be H_i = (- {\bf A}^{-1} {\bf B})^i H_0 \ee
and
\be H = \sum_i H_i = {1 \over 1 + {\bf A}^{-1} {\bf B}} H_0 \ .\label{sum}\ee
That such a formal expression for the soluton is acceptable is
predicated on the fact that the operator ${\bf A}$ is separable and
therefore invertible. In fact, one can show that the differential
equation \be \left( (1 - z^2) \partial_z^2 - 2 z \partial_z +
n(n+1)\right) f(z) = 0 \ee is solved by
\be f(z) = L_n(z) \ee
where $L_n(z)$ is the Legendre polynomial of degree $n$.  This is the
natural basis to work in when acting with ${\bf A}^{-1}$. To generate
the recursive sum, one must act with ${\bf B}$, expand the $z$
dependence in Legendre polynomial basis, and convolve the Green's
function with respect to $\rho_1$ and $\rho_2$. As a proof of
principle, we will compute the first few terms in the expansion.

An effective technique for computing the action of ${\bf A}^{-1}$ is
the method of undetermined coefficients.  Acting with ${\bf B}$, we
find that
\beq
{\bf B}H_0 =\frac{96 z \rho_1^2\rho_2^2}{k^2R_2^2(\rho_1^2+\rho_2^2)^5}
-\frac{24(\rho_1^2-\rho_2^2)^2}{k^2R_2^2(\rho_1^2+\rho_2^2)^5}.
\label{BH0}
\eeq
We then consider a general linear combination of basis functions for
which action by ${\bf A}$ produces terms of the form in (\ref{BH0}).
The basis functions must satisfy the following properties.  First,
they may only depend on $\rho_i$ through $\rho_i^2$, and must be
symmetric under interchanging $\rho_1$ and $\rho_2$. Second, for
physical reasons we expect poles only of the form
$\frac{1}{(\rho_1^2 +\rho_2^2)^n}$. Third, $H_1$ should not contain
any factors more divergent than $\frac{1}{(\rho_1^2 +\rho_2^2)^4}$.
Once the power of $\frac{1}{(\rho_1^2 +\rho_2^2)}$ is established,
there will be an additional coefficient of $\rho_1^2\rho_2^2$ as
determined by dimensional analysis (up to a change of basis
functions.)  This motivates the ansatz
\beq
k^2R_2^2 H_1 = c_1 \frac{z \rho_1^2\rho_2^2}{(\rho_1^2+\rho_2^2)^4}
+c_2
\frac{z}{(\rho_1^2+\rho_2^2)^2} +c_3 \frac{
\rho_1^2\rho_2^2}{(\rho_1^2+\rho_2^2)^4}+c_4\frac{1}{(\rho_1^2+\rho_2^2)^2}
\eeq
which solves the first stage of the recursion relation for
$c_1=2,c_2=0,c_3=2,c_4=-1$, or
\be
H_1=\frac{1}{k^2R_2^2}\left(\frac{2(1+z)
\rho_1^2\rho_2^2}{(\rho_1^2+\rho_2^2)^4}-\frac{1}{(\rho_1^2+\rho_2^2)^2}\right) \ . \ee
A similar calculation produces for $H_2$:
\beq
H_2=\frac{1}{k^4R_2^4}\left(\frac{8}{3}\frac{(1+z)^2
\rho_1^4\rho_2^4}{(\rho_1^2+\rho_2^2)^5}-\frac{26}{9}\frac{(1+z)\rho_1^2\rho_2^2}{(\rho_1^2+\rho_2^2)^3}
+\frac{10}{9}\frac{1}{(\rho_1^2+\rho_2^2)}\right) \ .
\eeq
At each order in the recursion, there are finitely many basis
functions, so this method can be applied at any order. It quickly
becomes clear, though, that at higher orders the explicit
calculations become quite cumbersome.
Nevertheless, $H_1$ and $H_2$ do appear to have some pattern
suggesting that perhaps there is some way to resum this series.

By substituting this solution into the ansatz (\ref{ansatz}), and
scaling, as usual for the M2-branes \cite{Maldacena:1997re},
\be \rho_1 = l_p^{3/2} U_1^{1/2}, \qquad \rho_2 = l_p^{3/2} U_2^{1/2} \ , \ee
we will obtain a supergravity dual of the decoupled field theory. To
see the structure of this solution, let us first examine the scaling
of the metric of ${\cal M}_8$
\be ds_{{\cal M}8}^2 = l_p^3 dS_{{\cal M}8} ^2(U_1,\theta_1,\phi_1,\varphi_1,U_2,\theta_2,\phi_2,\varphi_2) \ee
where
\beq dS_{{\cal M}8}^2 &=&
\left(1+ {U_1 \over c k^2 g_{YM2}^2}\right)
\left({1 \over 4 U_1 } d U_1^2  + {U_1 \over 4} (d \theta_1^2 + \sin^2 \theta_1 d \phi_1^2) \right)
\cr
&&+ \left(1 + {U_2 \over k^2 c g_{YM2}^2}\right)
\left({1 \over 4 U_2} d U_2^2  + {U_2 \over 4} (d \theta_2^2 + \sin^2 \theta_2 d \phi_2^2) \right) \cr
&& -{U_1 U_2 \over 2 c g_{YM2}^2 k^2} \left( {d \vec r_1 \cdot d \vec r_2 \over r_1 r_2} \right) \cr
&&+ \left( \begin{array}{cc}
  {1 \over U_1} + {1 \over  U_2} & {k \over  U_2} \label{dS}\\
{k \over U_2} & {1 \over c g_{YM2}^2} + {k^2 \over U_2}\end{array}\right)^{-1\, ij} (d \varphi_i + A_i) (d \varphi_j + A_j)
\eeq
has the dimension of inverse length and is independent of $l_p$. We
have expressed $R_2$ in terms of the field theory parameter
\be R_2^2= c g_{YM2}^2 l_p^3 \ee
by combining (\ref{scaling}) with the standard relation
\be l_p = g^{1/3} l_s \ . \ee

Let us also introduce a scaled harmonic function
\be h(U_1,\theta_1,\phi_1,U_2,\theta_2,\phi_2) = l_p^3 H \ee
which is also independent of $l_p$. Using  (\ref{seed}) as $H_0$, we have
\beq h_0 &=& l_p^3 H_0 = {c' N \over (U_1 + U_2)^3} \cr
h_1 & = & l_p^3 H_1 = {c' N \over c g_{YM2}^2 k^2} \left( {2 (1+z) U_1 U_2 \over (U_1+U^2)^4} - {1 \over (U_1+U_2)^2}\right) \label{h123}\\
h_2 & = &  l_p^3 H_2 = {c' N \over c^2 g_{YM}^4 k^4}  \left( {8
(1+z)^2 U_1^2 U_2^2  \over 3 (U_1+U_2)^5} - {26 (1+z) U_1 U_2  \over
9 (U_1+U_2)^3}+{10 \over 9 (U_1+U_2)} \right)
\nonumber \eeq
which indeed is independent of $l_p$. In terms of these quantities,
the supergravity solution we are after takes the form
\be ds^2 = l_p^2 \left[h^{-2/3} (-dt^2+dx_1^2+dx_2^2) + h^{1/3} dS_{{\cal M}8}^2 \right] \label{final} \ee
where the expression inside the square bracket only depends on field
theory variables and not on $l_p$. Also, for small $U_1$ and $U_2$,
the geometry asymptotes to $AdS_4 \times S_7 / Z_k$. It is also
straight forward to reduce this geometry to type IIA. These
geometries capture the renormalization group flow of
Chern-Simons/Yang-Mills/matter system down to ${\cal N}=6,8$
superconformal Chern-Simons/matter theory, and is effective for $N
\gg k$.

The explicit solution to the eleven dimensional supergravity
equations of motion given in (\ref{sum}), (\ref{dS}), (\ref{h123}),
and (\ref{final}) is the main result of this paper. Admittedly, the
solution we found is not in an ideal form.  The recursive nature of
the solution presented here makes it cumbersome to evaluate and
display the function even using numerical methods. Still, the form
that the solution takes for large and small $U_i$ was clear from the
beginning. The recursive procedure provides the details of the
solution near the cross-over region at the scale $g_{YM2}^2 k$
corresponding to the mass deformation due to the Chern-Simons term.

The eight dimensional hyper-K\"ahler geometry we studied in this paper
has quite a bit of structure \cite{Gauntlett:1997pk}. The fact that a
recursive procedure for solving for the Greens function on this space
suggests the possibility that there exists more elegant approaches to
the problem we considered. Green's functions in self-dual four
manifolds have been analyzed using various methods
\cite{Page:1979ga,Atiyah:1981ey}. Perhaps some of these methods can be
applied to the problem considered in this paper.

\section*{Acknowledgements}

We would like to thank
O. Aharony, O. Bergman, S. Cherkis, S. Hirano, N. Itzhaki, and J.
Sonnenschein for discussions. AH is grateful to Tel Aviv University
and Weizmann Institute where part of this work was done. This work
was supported in part by the DOE grant DE-FG02-95ER40896, NSF CAREER
Award No. PHY-0348093, a Cottrell Scholar Award from Research
Corporation, and funds from the University of Wisconsin.

\bibliography{hk}\bibliographystyle{utphys}

\end{document}